\documentclass[aip,cha,preprint,numerical]{revtex4-1}
\usepackage{graphicx}
\usepackage{amsmath}
\usepackage{amssymb}
\usepackage{psfrag}

\newcommand{\dif}{\mathrm{d}}

\setcitestyle{numbers,open={[},close={]}}

\begin{document}
%------------------------------------------------------------------------------%
\title{Scaling of energy spreading in a disordered Ding-Dong lattice}

\author{A. Pikovsky}
\affiliation{Institute for Physics and Astronomy, University of Potsdam, 
Karl-Liebknecht-Str. 24/25, 14476 Potsdam-Golm, Germany.
}
\affiliation{Department of Control Theory, Nizhny Novgorod State University, 
Gagarin Av.~23, Nizhny Novgorod, 603950, Russia.}

\begin{abstract}
We study numerically propagation of energy in a one dimensional Ding-Ding lattice,
composed of linear oscillators with ellastic collisions.
Wave propagation is suppressed by breaking translational symmetry, we consider 
three way to do this: a position disorder, a mass disorder, and a dimer lattice with alternating distances
between the units. In all cases the spreading of an initially localized wavepacket is irregular, due
to appearance of chaos, and subdiffusive. Guided by a nonlinear diffusion equation, we establish
that the mean waiting times of spreading obey a scaling law in dependence on energy. Moreover,
we show that the spreading exponents very weakly depend on the level of disorder.
\end{abstract}
%\date{\today}
%\pacs{}
%\date{\today}

\maketitle
\section{Introduction}
\label{sec:intro}
Nonequilibrium processes in nonlinear Hamiltonian lattices are a subject of intensive research.
Here typically three setups are considered. In one setting, thermal conductivity of such a lattice
is of interest, thus one couples it to two thermostates with different temperatures and studies
the properties of the energy flux;  an early review of these studies is in Ref.~\cite{Lepri-Livi-Politi-03},
for recent progress see~\cite{Mejia-Monasterio_etal-19,Giberti-19}. Another setting deals with properties
of a relaxation toward an equilibrium chaotic state with constant energy density. Here one identifies
modes which can be treated as the first and second sound~\cite{Gendelman-Savin-10,Gendelman_etal-12,Pikovsky-15a}. 
Some of these modes can be described theoretically in the framework of 
fluctuating hydrodynamics~\cite{Beijeren-12,Mendl-Spohn-13}.
In this paper we follow the third widely used setup, where a spreading of an initially localized wavepacket
into vacuum is considered. In a regular lattice, such a spreading is of course dominated by linear or nonlinear waves,
propagating with a constant speed.
However, in a disordered lattice, due to Anderson localization, there are no propagating linear waves and the
spreading typically appears as a slow subdiffusion. A very popular model here is DANSE, discrete Anderson
nonlinear Schroedinger
lattice~\cite{Pikovsky-Shepelyansky-08,Fishman-Krivolapov-Soffer-12,Lapteva_etal-14}. Another class of
models are oscillator chains~\cite{Skokos_etal-09,Mulansky-Ahnert-Pikovsky-11,Mulansky-Pikovsky-12a,Achilleos-18},
in contradistinction to Schroedinger lattices they posses only one integral of motion (energy) and therefore are
simpler to treat. Spreading of energy in such systems is due to chaos, which in course of spreading and the corresponding
decrease of the energy density becomes 
weaker~\cite{Mulansky-Ahnert-Pikovsky-Shepelyansky-11,Pikovsky-Fishman-11,Senyange_etal-18,Ngapasare_etal-19}.
This leads to the slowing of the spreading, and there are suggestions that the spreading may 
eventually stop~\cite{Johansson-Kopidakis-Aubry-10}. This still unsolved puzzle makes further numerical investigations
of disordered nonlinear lattices, especially at very large times, extremely relevant.

In this paper we address the problem of energy spreading for a Ding-Dong 
model~\cite{Prosen-Robnik-92,Roy-Pikovsky-12}, which has several nice properties.
The model is formulated as a chain of linear oscillators, interacting via elastic collisions (see Section~\ref{sec:Model}
below). Thus, calculation of its evolution in time is quite simple, because one can write an analytic solution between
the collision events and calculate next collision times. This allows  following the evolution until
very large times without essential loss of accuracy. In our calculations 
below the maximal times are $10^{9}-10^{10}$, to be compared
with the characteristic period of oscillators $2\pi$. Furthermore, the Ding-Dong model belongs to a class of 
strongly nonlinear 
lattices~\cite{Nesterenko-83,Rosenau-Hyman-93,Rosenau-Pikovsky-05,Pikovsky-Rosenau-06,Ahnert-Pikovsky-09,Mulansky-Pikovsky-13},
which are characterized by ``sonic vacuum'': linear modes (even localized ones like in disordered models)
do not exist. This makes the edges of the spreading wavepacket extremely (superexponentially
for lattices with a smooth potential) sharp
and allows for characterization of propagation via the edge velocity~\cite{Mulansky-Pikovsky-13}. Below we define
the Ding-Dong model and different types of disorder in it in Section~\ref{sec:Model}.
In Section \ref{sec:scal} main scaling characteristics  of the spreading of energy are introduced, which are 
numerically explored in different configuration in Section~\ref{sec:nr}.
Finally, we discuss the results and compare them with properties of other
strongly nonlinear lattices in Section~\ref{sec:concl}.

%------------------------------------------------------------------------------%
\section{Disordered Ding-Dong lattice}\label{sec:Model}
%------------------------------------------------------------------------------%
The Ding-Dong lattice~\cite{Prosen-Robnik-92} is a chain of linear oscillators described by
Hamiltonian
\begin{equation}
H=\sum_k \left(\frac{p_k^2}{2M_k}+M_k\frac{q_k^2}{2}\right)\;.
\label{eq:dd1}
\end{equation}
Masses of oscillators $M_k$ are generally different, but all the oscillators have the same frequency $\omega=1$.
The oscillators are aligned along a line, with generally different distances $R_{k,k+1}$ between them.
The interaction of oscillators is due to elastic collisions: when oscillators $k$ and $k+1$ collide, i.e. when
$q_k=R_{k,k+1}+q_{k+1}$, they exchange their momenta according to
\begin{equation}
  p_k\to \frac{2M_{k}p_{k+1}+(M_k-M_{k+1})p_k}{M_k+M_{k+1}}\;,\qquad
 p_{k+1}\to\frac{2M_{k+1}p_k-(M_k-M_{k+1})p_{k+1}}{M_k+M_{k+1}}\;.
\label{eq:massdis}
\end{equation}

The Ding-Dong model belongs to a class of strongly nonlinear lattices: here in the linear approximation
of infinitesimal oscillation amplitudes, any interaction between units is absent, so no waves are 
propagating (so-called sonic vacuum). In a regular Ding-Dong lattice, i.e. when $R_{k,k+1}=R=const$ 
and $M_k=M=const$, nonlinear waves, compactons, can propagate~\cite{Prosen-Robnik-92,Roy-Pikovsky-12}.
As we want to avoid compactons in this study, we consider three types of inhomogeneous Ding-Dong lattices:
\begin{enumerate}
\item \textit{Position-disorder lattice:} here we assume that all the masses are the same $M_k=1$, but
the distances between the oscillators are distributed randomly. Namely, the distances are chosen in each realization
of disorder as independent random numbers from a uniform distribution $1-r<R<1+r$.
\item \textit{Mass-disorder lattice:} here we assume that all the distances are the same $R_{k,k+1}=1$, but the
masses are chosen as independent random numbers with a  uniform distribution $1-m<M<1+m$.
\item \textit{Dimer lattice:} here we break the homogeneity in the simplest way, assuming alternating distances
between the oscillators: $R_{k,k+1}=1+d(-1)^k$, while all the masses are equal $M_k=1$.
\end{enumerate}

\begin{figure}[h!]
 \psfrag{xlabel}[cc][cc]{size $X$}
  \psfrag{ylabel2}[cc][cc]{$\Delta T$}
  \psfrag{ylabel1}[cc][cc]{$T$}
 	\centering
 	\includegraphics[width=\columnwidth]{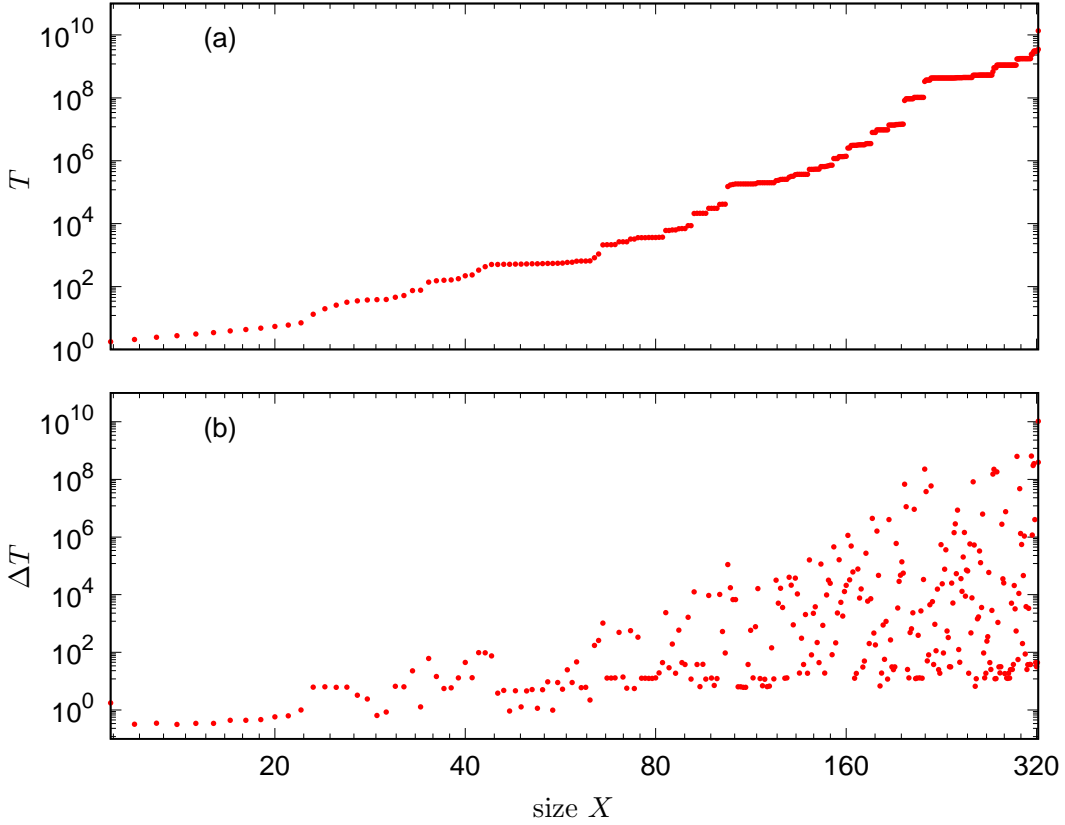}
\caption{Example of spreading of an initially local perturbation with energy $E=10$ in a position-disorder lattice with $r=0.1$.
Upper panel: time $T(X)$, at which the active region achieves size $X$. Bottom panel: waiting times $\Delta T(X)$. 
 }
\label{fig:espr}
\end{figure}

In all these cases we consider the problem of energy spreading from a local in space initial distribution
of energy. We fix the initial energy $E$ (which is, of course, conserved in course of evolution), and prepare
initial conditions by setting initial momenta of 10 neighboring particles to be nonzero, all initial coordinates are zero:
\begin{equation}
q_k(0)=0,\qquad p_k(0)=\begin{cases} S N(0,1)& \text{ for } 1\leq k\leq 10\;,\\
0 &\text{else}\;.
\end{cases}
\label{eq:inc}
\end{equation}
Here $N(0,1)$ are Gaussian random numbers, and $S$ is a normalization factor ensuring $\sum_k \frac{p_k^2(0)}{2M_k}=E$.
Then we run the evolution, and observed an irregular spreading of energy. It is illustrated in Fig.~\ref{fig:espr}, were we show
an example of this spreading. At each moment of time, there is an ``excited domain'' $k_{-}\leq k \leq k_{+}$ where 
the energies of oscillators
are positive, while outside of this domain all the oscillators are at rest. The size of the excited domain $X=k_+-k_-+1$
extends by one at
certain moments of time $T(X)$, when a neighboring oscillator at rest is hit by an active one. In Fig.~\ref{fig:espr}
we plot these times as a function of the corresponding seizes $T(X)$, additionally we plot ``waiting times''
$\Delta T(X)=T(X+1)-T(X)$ which show, how long it takes to extend the size of the excited domain by one unit.
The calculations in Fig.~\ref{fig:espr} ended when the waiting time exceeded $10^{10}$.

\section{Scaling of energy spreading}
\label{sec:scal}
The main goal of our study is to find statistical properties of the energy spreading, qualitatively presented in Fig.~\ref{fig:espr}.
Similarly to the analysis of strongly nonlinear lattices with a smooth potential~\cite{Mulansky-Pikovsky-13},
we use a nonlinear diffusion equation (NDE) as a guiding phenomenological tool.
The NDE equation, where the variable $\rho(x,t)$ should be interpreted as an energy density, reads
\begin{equation}  \label{eqn:NDE}
 \frac{\partial \rho}{\partial t} = D_0 \frac\partial{\partial x} \left(\rho^a \frac{\partial \rho}{\partial x}\right) = \frac{D_0}{a+1} \frac{\partial^2}{\partial x^2} \rho^{a+1},\qquad \mbox{with} 
 \qquad \int \rho\, \dif x = E\;.
\end{equation} 
This equation posseses a self-similar solution, describing a spreading domain
\begin{equation} \label{eqn:self_sim_solution}
\rho(x,t) = 
\left\{
\begin{array}{ll}
  (t-t_0)^{-1/(2+a)} \left(c\,E^{2a/(a+2)}-\frac{a x^2}{2(a+2)(t-t_0)^{2/(2+a)}}\right)^{1/a} & \mbox{for} \quad |x|<x_m\;,\\
    0 & \mbox{for} \quad |x|>x_m\;.
\end{array}
\right.
\end{equation}
Here $c$ is a constant, and $x_m$ denotes the edge of the ``excited domain''
\begin{equation} \label{eqn:edge_propagation}
  x_m = \sqrt{2c\frac{2+a}a} E^{a/(2+a)}(t-t_0)^{1/(2+a)}\;.
\end{equation}
This quantity has the same meaning as $X$ for the Ding-Dong lattice, so our aim is to use
\eqref{eqn:edge_propagation} for finding scaling of the spreading presented in Fig.~\ref{fig:espr}.
Here it is convenient to eliminate an additional parameter $t_0$, according to \cite{Mulansky-Pikovsky-13}
this can be accomplished by calculating the time derivative of the edge propagation:
\begin{equation} \label{eqn:vel}
 \frac{1}{E}\frac{\mbox{d} t}{\mbox{d} x_m}  \sim \left(\frac{E}{x_m}\right)^{-a-1}\;.
\end{equation}
In comparing with numerics for the Ding-Dong lattice, we have to identify $x_m$ with $X$,
and the inverse velocity of the propagation $\frac{\mbox{d} t}{\mbox{d} x_m} $ with the waiting 
time $\Delta T$. Because of strong fluctuations in particular realizations (cf. Fig.~\ref{fig:espr}),
we average $\log_{10}\Delta T$ in narrow ranges around exponentially spaced values of $X$. 
As a result, we obtain
dependencies $\log_{10}\Delta T$ vs $X$ for different values of the energy $E$ and different values 
of disorder. Scaling relation~\eqref{eqn:vel} predicts, that graphs of $\log_{10}\Delta T/E$ vs $E/X$
should collapse and provide a power-law dependence with some exponent $a$.

\section{Numerical results}
\label{sec:nr}
\subsection{Scaling of the wavepacket spreading}
As outlined above, we performed a series of simulations of spreading in disordered Ding-Dong
lattices, here we present the obtained statistical results.

\begin{figure}[h!]
\psfrag{xlabel1}[cc][cc]{$\log_{10}X$}
\psfrag{xlabel2}[cc][cc]{$\log_{10}(X/E)$}
\psfrag{ylabel1}[cc][cc]{$\log_{10}\Delta T$}
\psfrag{ylabel2}[cc][cc]{$\log_{10}\Delta T/E$}
 	\centering
 	\includegraphics[width=\columnwidth]{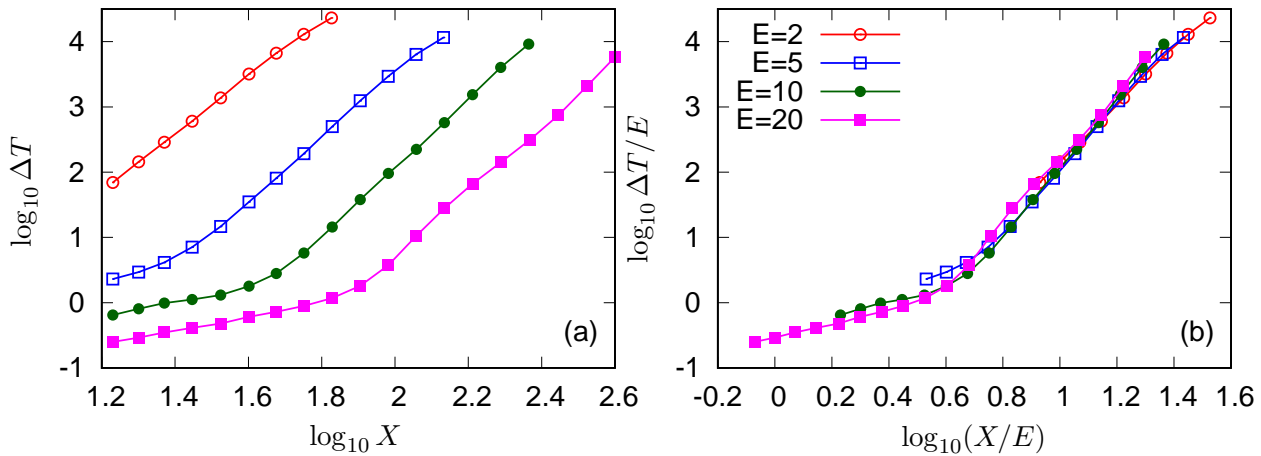}
\caption{Panel (a): averaged waiting times in a lattice with position disorder $r=0.1$, and four values of initial energies.
Panel (b): the same data in the scaled coordinates.
 }
\label{fig:scal1}
\end{figure}

Figure~\ref{fig:scal1} illustrates the scaling procedure we apply to characterize the spreading.
Panel (a) shows spreading in a lattice with distance disorder $r=0.15$, for four different values of
the energy $E$. The same data are shown in panel (b) in the scaled form. One can see a rather good collapse of the 
data, just conforming the qualitative prediction of the NDE model.

\begin{figure}[h!]
\psfrag{xlabel1}[cc][cc]{$\log_{10}(X/E)$}
\psfrag{ylabel1}[cc][cc]{$\log_{10}\Delta T/E$}
 	\centering
 	\includegraphics[width=\columnwidth]{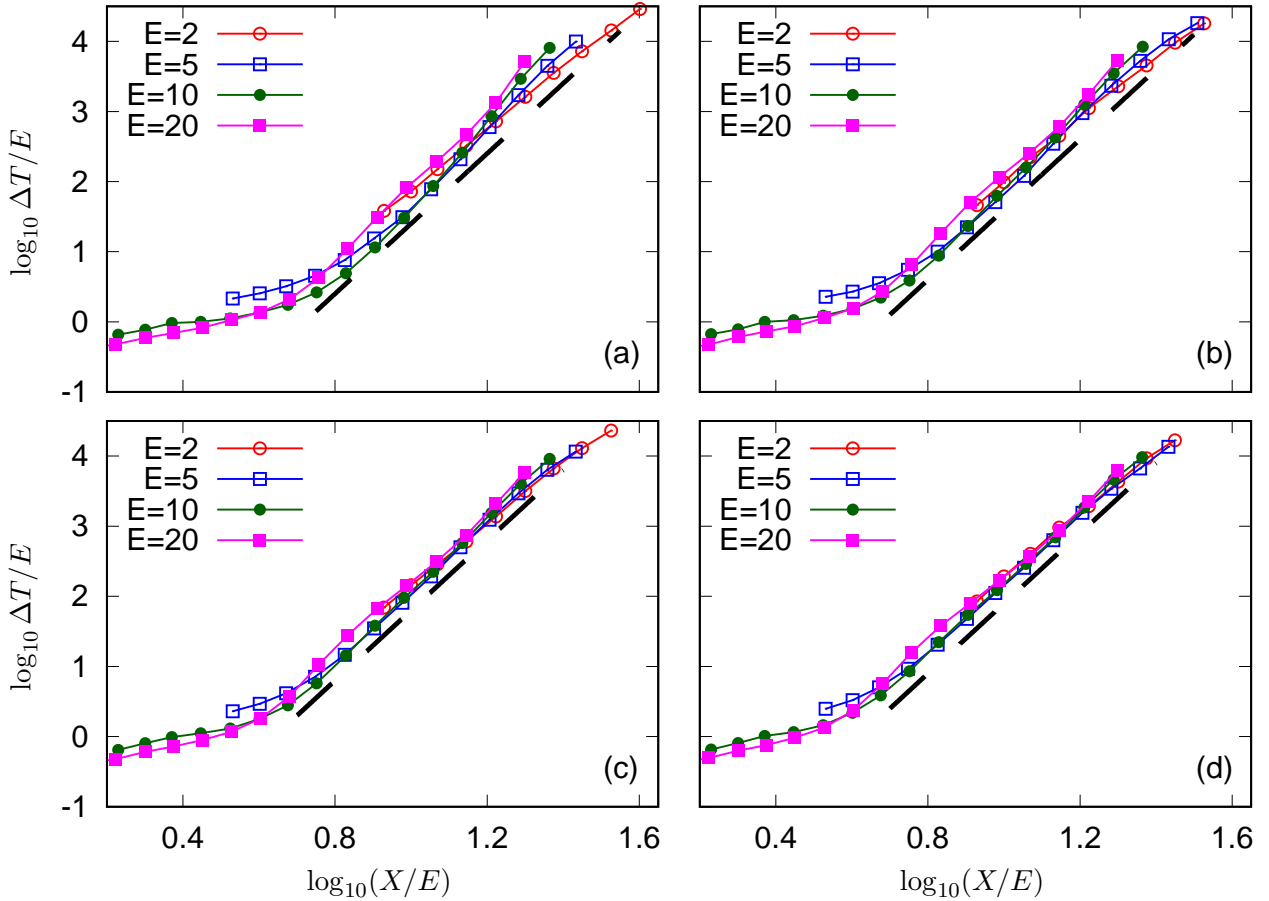}
\caption{Spreading in lattices with distance disorder, $r=0.05;\;0.1;\;0.15;\;0.2$ in panels (a-d), correspondingly.
The dashed black lines have slope 5.
 }
\label{fig:scal2}
\end{figure}

In Fig.~\ref{fig:scal2} we show the same scaling as in Fig.~\ref{fig:scal1}, but for four different
levels of disorder: in (a) $r=0.05$, in (b) $r=0.1$, in (c) $r=0.15$, and in (d) $r=0.2$. One can see
that there are two clearly different regions of scaling, with a crossover at the energy density $\approx 0.25$
(i.e. at $\log_{10}(X/E)\approx 0.6$). At larger densities the spreading is rather fast and the waiting time slowly 
grows as the density decreases. For smaller densities, there is a very strong power-law dependence
of the waiting time on density
\begin{equation}
\Delta T \sim E\left(\frac{E}{X}\right)^{-\alpha_{pd}},\quad \alpha_{pd}\approx 5\;.
\label{eq:sc5}
\end{equation}
Equivalently, one can express this as a spreading law
\begin{equation}
X\sim T^{1/(\alpha_{dd}+1)}\approx T^{1/6}\;.
\label{eq:sc5-1}
\end{equation}

\begin{figure}[h!]
\psfrag{xlabel1}[cc][cc]{$\log_{10}(X/E)$}
\psfrag{ylabel1}[cc][cc]{$\log_{10}\Delta T$}
 	\centering
 	\includegraphics[width=\columnwidth]{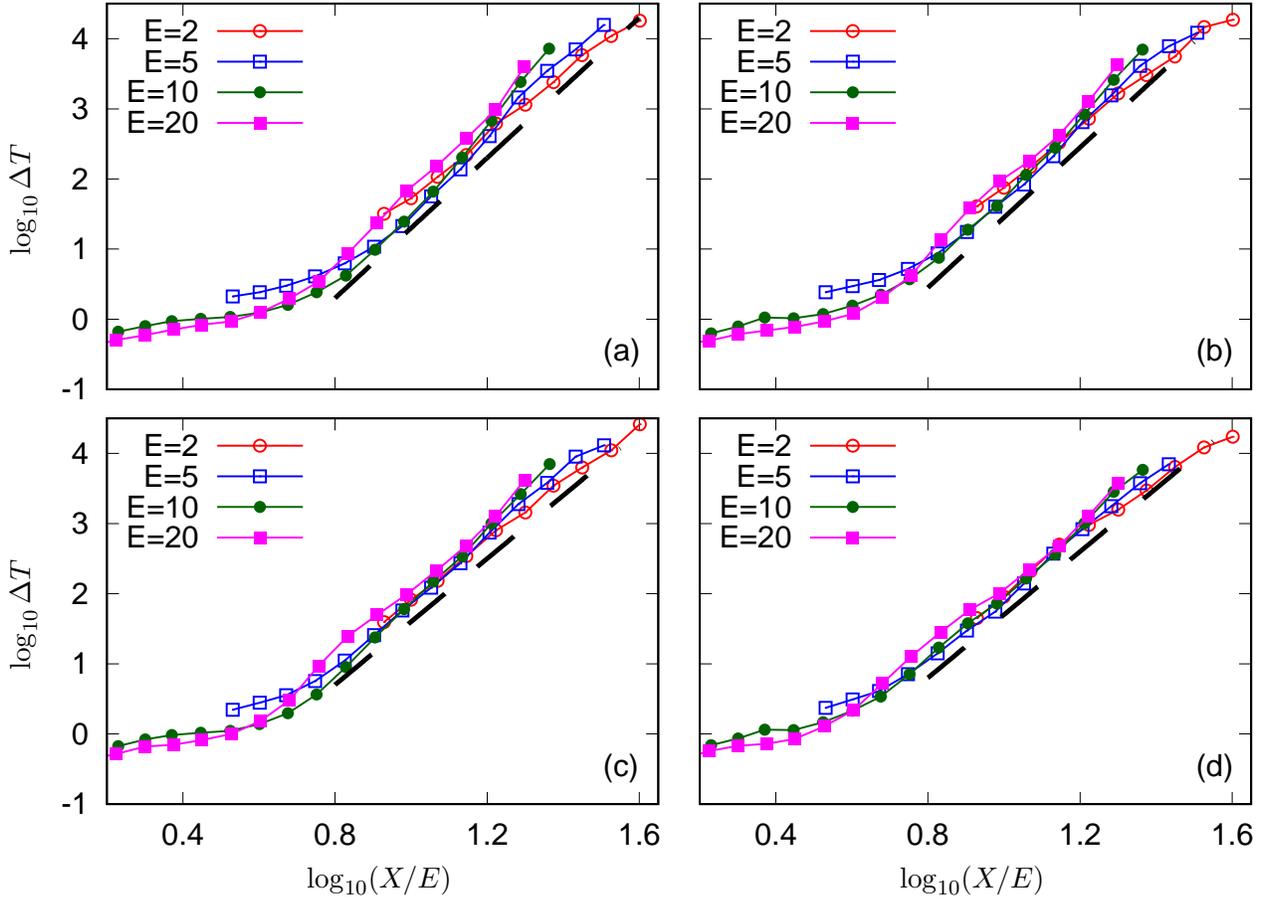}
\caption{The same as Fig.~\ref{fig:scal2}, but for mass disorder, $m=0.05;\;0.1;\;0.15;\;0.2$ in panels (a-d), respectively.
The dashed black lines have slope $5$ in panels (a,b) and $4.5$ in panels (c,d).
 }
\label{fig:scal3}
\end{figure}

\begin{figure}[h!]
\psfrag{xlabel1}[cc][cc]{$\log_{10}(X/E)$}
\psfrag{ylabel1}[cc][cc]{$\log_{10}\Delta T$}
 	\centering
 	\includegraphics[width=\columnwidth]{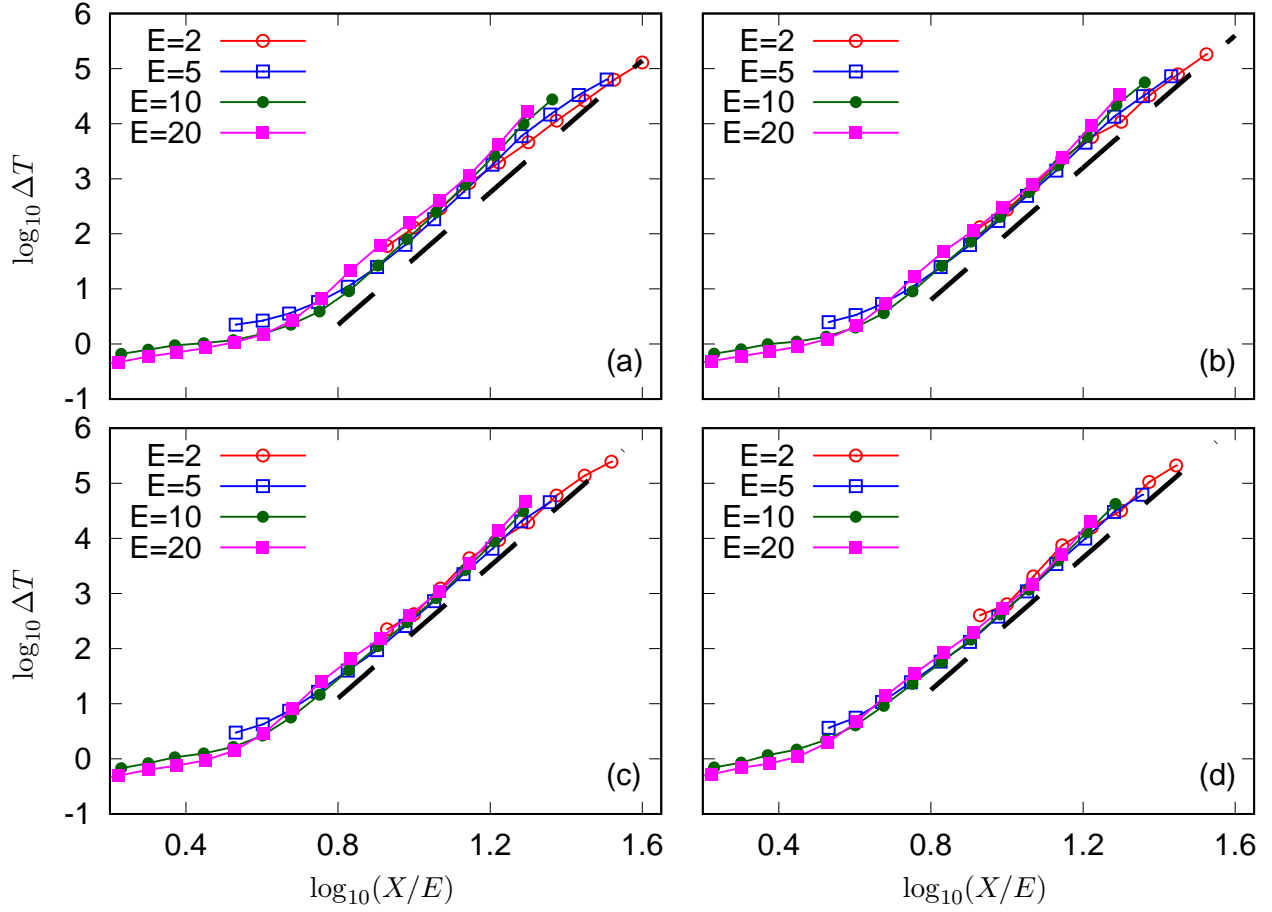}
\caption{The same as Fig.~\ref{fig:scal2}, but for dimer lattices, with $d=0.05;\;0.1;\;0.15;\;0.2$ in panels (a-d).
The dashed black lines have slope $6$.
 }
\label{fig:scal4}
\end{figure}

Next, we present a similar scaling analysis for the Ding-Ding lattice with mass disorder (Fig.~\ref{fig:scal3}),
and for dimer lattice (Fig.~\ref{fig:scal4}). Qualitatively, all the results are similar, however with small 
quantitative differences.
For the dimer lattice (Fig.~\ref{fig:scal4}),  for all values of the ``disorder'' $d$ the scaling exponent
in \eqref {eq:sc5} is $\alpha_{dim}\approx 6$. For the random masses lattice (Fig.~\ref{fig:scal3}),
for small disorders $m=0.05,0.1$ the best fit is with exponent  $\alpha_{m}\approx 5$, while for large disorder
$m=0.15,0.2$ the value of the exponent is different $\alpha_m\approx 4.5$.

\subsection{Distribution of waiting times}
\begin{figure}[h!]
\psfrag{xlabel1}[cc][cc]{$\log_{10}(\tau)$}
\psfrag{ylabel1}[cc][cc]{$\log_{10}P(\tau)$}
\psfrag{dis}[cc][cc]{$r$}
 	\centering
 	\includegraphics[width=0.6\columnwidth]{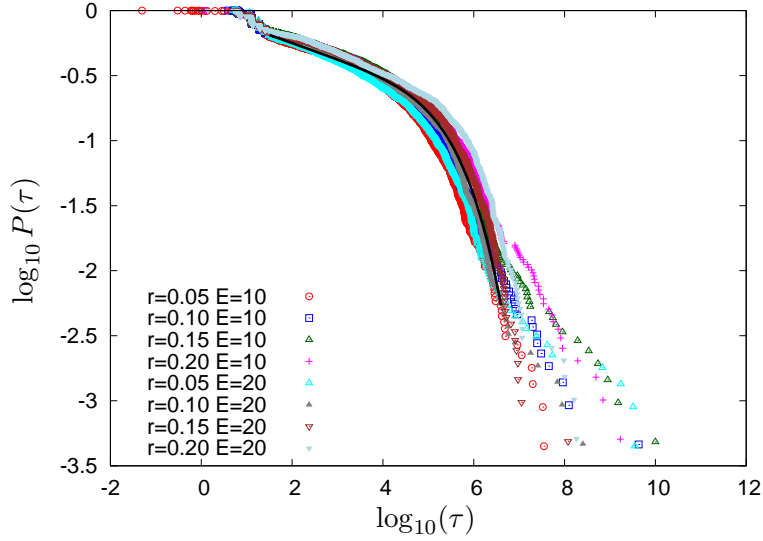}
\caption{Distribution of waiting times for the lattice with
different values of position disorder, for two values of the total energy, at a predescribed level of the average density. 
Relatively good overlap
of all curves indicates that dependence on disorder is very small.
 }
\label{fig:td_dd}
\end{figure}

Here we present results on the distribution of the waiting times. 
It is convenient to calculate the cumulative distribution $P(\tau)=\text{prob}(\Delta T >\tau)$. 
In Fig.~\ref{fig:td_dd} we show these distributions for the case of position disorder. In
order to compare different energies, we considered waiting times at the same energy density.
Namely, in Fig.~\ref{fig:td_dd} we show for energy $E=20$ the distribution of waiting times at
the wavepacket size $X=300$, and for $E=10$ the same quantity at $X=150$. One can see a good overlap
of all curves, what means that waiting times only weakly depend on the level of disorder.
The solid line shows an empirical fit of the distribution as a power-law with a stretched-exponential cutoff:
$P(\tau)\sim \tau^{-c}10^{-a\tau^b}$ with $c=1/8$, $a=4\cdot 10^{-7}$, $b=8$. 

\begin{figure}[h!]
\psfrag{xlabel1}[cc][cc]{$\log_{10}(\tau d^{-3/2})$}
\psfrag{ylabel1}[cc][cc]{$\log_{10}P(\tau)$}
\psfrag{dis}[cc][cc]{$r$}
 	\centering
 	\includegraphics[width=0.6\columnwidth]{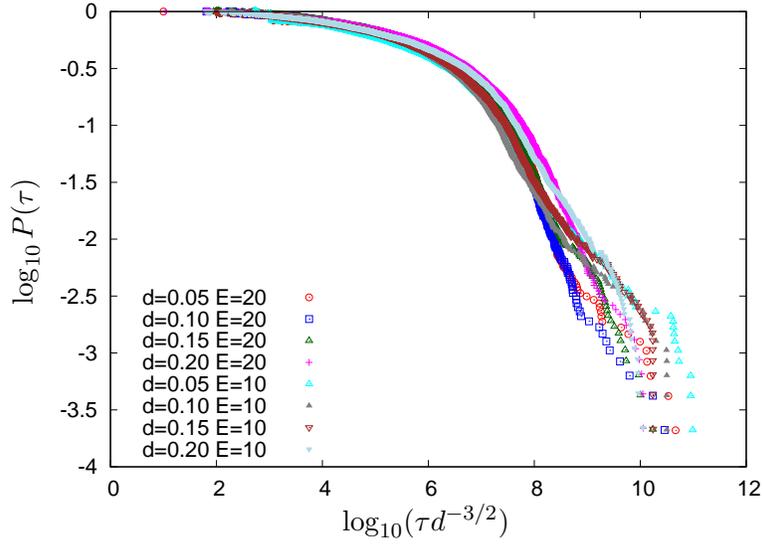}
\caption{The same as Fig.~\ref{fig:td_dd}, but for dimer lattices. Here a good overlap is obtained
when the time is rescaled with disorder level, as indicated. 
 }
\label{fig:td_di}
\end{figure}

A different property shows the dimer lattice: here the average waiting time at a certain density
significantly depends on disorder.
We illustrate this in Fig.~\ref{fig:td_di}, which is quite similar to Fig.~\ref{fig:td_dd}, but here to achieve an overlap
of different distributions, a rescaled time is used. We have found that characteristic waiting time scales with
disorder like
$\tau \sim d^{3/2}$. Furthermore, in the dependence $P(\tau)$ one cannot recognize a region with a nearly power-law
profile, contrary to the case of distance disorder Fig.~\ref{fig:td_dd}.

\section{Discussion}
In this paper we performed a comprehensive numerical exploration
of energy spreading in the disordered Ding-Dong model, for different types and strengths of disorder, and
for different initial energies. First, we have demonstrated, that for all parameters, the scaling \eqref{eqn:vel},
corresponding to a nonlinear diffusion equation, works well. Moreover, the characteristic exponent
$a$ in \eqref{eqn:vel} appears to have very similar values for all cases with mass and position disorder, and a larger
value for the dimer lattice. This power-law dependence of waiting times means that  the spreading at large 
times is  subdiffusive, $X\sim T^{1/6}$ for distance and mass disorder, and $X\sim T^{1/7}$ for dimer lattice.

It is instructive to compare these findings with the results for lattices composed of 
linear oscillators coupled by higher-order smooth
potentials~\cite{Mulansky-Pikovsky-13}. In the latter case the scaling  \eqref{eqn:vel} works as well,
however the resulting curve is not a straight line in log-log coordinates, 
what means that a single parameter
$a$ does not exist, rather it increases in course of spreading. Correspondingly, the spreading 
is slower than subdiffusion.

We have also found that disorder has only a small effect on the propagation speed, only for dimer lattice
the value of position modulation influences the characteristic waiting time; for lattices with distance and mass
disorder this effect is very small. On the other hand, we have not studied lattices with very strong 
(beyond 20\% modulation) disorder.

Finally, we would like to mention that a two-dimensional generalization of the Ding-Dong model
is possible; this problem is under investigation now.
\label{sec:concl}

\acknowledgments
The works was supported by the Russian Science Foundation (Grant 17-12-01534).


\begin{thebibliography}{10}%
\makeatletter
\providecommand \@ifxundefined [1]{%
 \ifx #1\undefined \expandafter \@firstoftwo
 \else \expandafter \@secondoftwo
\fi
}%
\providecommand \@ifnum [1]{%
 \ifnum #1\expandafter \@firstoftwo
 \else \expandafter \@secondoftwo
\fi
}%
\providecommand \enquote [1]{``#1''}%
\providecommand \bibnamefont  [1]{#1}%
\providecommand \bibfnamefont [1]{#1}%
\providecommand \citenamefont [1]{#1}%
\providecommand\href[0]{\@sanitize\@href}%
\providecommand\@href[1]{\endgroup\@@startlink{#1}\endgroup\@@href}%
\providecommand\@@href[1]{#1\@@endlink}%
\providecommand \@sanitize [0]{\begingroup\catcode`\&12\catcode`\#12\relax}%
\@ifxundefined \pdfoutput {\@firstoftwo}{%
 \@ifnum{\z@=\pdfoutput}{\@firstoftwo}{\@secondoftwo}%
}{%
 \providecommand\@@startlink[1]{\leavevmode}%
 \providecommand\@@endlink[0]{}%
}{%
 \providecommand\@@startlink[1]{%
  \leavevmode
  \pdfstartlink
   attr{/Border[0 0 1 ]/H/I/C[0 1 1]}%
   user{/Subtype/Link/A<</Type/Action/S/URI/URI(#1)>>}%
  \relax
 }%
 \providecommand\@@endlink[0]{\pdfendlink}%
}%
\providecommand \url  [0]{\begingroup\@sanitize \@url }%
\providecommand \@url [1]{\endgroup\@href {#1}{\urlprefix}}%
\providecommand \urlprefix [0]{URL }%
\providecommand \Eprint[0]{\href }%
\@ifxundefined \urlstyle {%
  \providecommand \doi [1]{doi:\discretionary{}{}{}#1}%
}{%
  \providecommand \doi [0]{doi:\discretionary{}{}{}\begingroup
  \urlstyle{rm}\Url }%
}%
\providecommand \doibase [0]{http://dx.doi.org/}%
\providecommand \Doi[1]{\href{\doibase#1}}%
\providecommand \selectlanguage [0]{\@gobble}%
\providecommand \bibinfo [0]{\@secondoftwo}%
\providecommand \bibfield [0]{\@secondoftwo}%
\providecommand \translation [1]{[#1]}%
\providecommand \BibitemOpen[0]{}%
\providecommand \bibitemStop [0]{}%
\providecommand \bibitemNoStop [0]{.\EOS\space}%
\providecommand \EOS [0]{\spacefactor3000\relax}%
\providecommand \BibitemShut [1]{\csname bibitem#1\endcsname}%
%</preamble>
\bibitem{Lepri-Livi-Politi-03}%
  \BibitemOpen
  \bibfield{author}{%
  \bibinfo {author} {\bibfnamefont{S.}~\bibnamefont{Lepri}}, \bibinfo {author}
  {\bibfnamefont{R.}~\bibnamefont{Livi}},\ and\ \bibinfo {author}
  {\bibfnamefont{A.}~\bibnamefont{Politi}},\ }%
  \bibfield{title}{%
  \enquote{\bibinfo {title} {Thermal conduction in classical low-dimensional
  lattices},}\ }%
  \bibfield{journal}{%
  \bibinfo {journal} {Physics Reports}\ }%
  \textbf{\bibinfo {volume} {377}},\ \bibinfo {pages} {1--80} (\bibinfo {year}
  {2003})\BibitemShut{NoStop}%
\bibitem{Mejia-Monasterio_etal-19}%
  \BibitemOpen
  \bibfield{author}{%
  \bibinfo {author} {\bibfnamefont{C.}~\bibnamefont{Mej\'{\i}a-Monasterio}},
  \bibinfo {author} {\bibfnamefont{A.}~\bibnamefont{Politi}},\ and\ \bibinfo
  {author} {\bibfnamefont{L.}~\bibnamefont{Rondoni}},\ }%
  \bibfield{title}{%
  \enquote{\bibinfo {title} {Heat flux in one-dimensional systems},}\ }%
  \bibfield{journal}{%
  \bibinfo {journal} {Phys. Rev. E}\ }%
  \textbf{\bibinfo {volume} {100}},\ \bibinfo {pages} {032139} (\bibinfo {year}
  {2019})\BibitemShut{NoStop}%
\bibitem{Giberti-19}%
  \BibitemOpen
  \bibfield{author}{%
  \bibinfo {author} {\bibfnamefont{C.}~\bibnamefont{Giberti}}, \bibinfo
  {author} {\bibfnamefont{L.}~\bibnamefont{Rondoni}},\ and\ \bibinfo {author}
  {\bibfnamefont{C.}~\bibnamefont{Vernia}},\ }%
  \bibfield{title}{%
  \enquote{\bibinfo {title} {Temperature and correlations in 1-dimensional
  systems},}\ }%
  \bibfield{journal}{%
  \bibinfo {journal} {The European Physical Journal Special Topics}\ }%
  \textbf{\bibinfo {volume} {228}},\ \bibinfo {pages} {129--142} (\bibinfo
  {year} {2019})\BibitemShut{NoStop}%
\bibitem{Gendelman-Savin-10}%
  \BibitemOpen
  \bibfield{author}{%
  \bibinfo {author} {\bibfnamefont{O.~V.}\ \bibnamefont{Gendelman}}\ and\
  \bibinfo {author} {\bibfnamefont{A.~V.}\ \bibnamefont{Savin}},\ }%
  \bibfield{title}{%
  \enquote{\bibinfo {title} {Nonstationary heat conduction in one-dimensional
  chains with conserved momentum},}\ }%
  \bibfield{journal}{%
  \bibinfo {journal} {Phys. Rev. E}\ }%
  \textbf{\bibinfo {volume} {81}},\ \bibinfo {pages} {020103} (\bibinfo {year}
  {2010})\BibitemShut{NoStop}%
\bibitem{Gendelman_etal-12}%
  \BibitemOpen
  \bibfield{author}{%
  \bibinfo {author} {\bibfnamefont{O.~V.}\ \bibnamefont{Gendelman}}, \bibinfo
  {author} {\bibfnamefont{R.}~\bibnamefont{Shvartsman}}, \bibinfo {author}
  {\bibfnamefont{B.}~\bibnamefont{Madar}},\ and\ \bibinfo {author}
  {\bibfnamefont{A.~V.}\ \bibnamefont{Savin}},\ }%
  \bibfield{title}{%
  \enquote{\bibinfo {title} {Nonstationary heat conduction in one-dimensional
  models with substrate potential},}\ }%
  \bibfield{journal}{%
  \bibinfo {journal} {Phys. Rev. E}\ }%
  \textbf{\bibinfo {volume} {85}},\ \bibinfo {pages} {011105} (\bibinfo {year}
  {2012})\BibitemShut{NoStop}%
\bibitem{Pikovsky-15a}%
  \BibitemOpen
  \bibfield{author}{%
  \bibinfo {author} {\bibfnamefont{A.}~\bibnamefont{Pikovsky}},\ }%
  \bibfield{title}{%
  \enquote{\bibinfo {title} {First and second sound in disordered strongly
  nonlinear lattices: numerical study},}\ }%
  \bibfield{journal}{%
  \bibinfo {journal} {J. Stat. Mech.}\ }%
  \textbf{\bibinfo {volume} {2015}},\ \bibinfo {pages} {P08007} (\bibinfo
  {year} {2015})\BibitemShut{NoStop}%
\bibitem{Beijeren-12}%
  \BibitemOpen
  \bibfield{author}{%
  \bibinfo {author} {\bibfnamefont{H.}~\bibnamefont{van Beijeren}},\ }%
  \bibfield{title}{%
  \enquote{\bibinfo {title} {Exact results for anomalous transport in
  one-dimensional hamiltonian systems},}\ }%
  \bibfield{journal}{%
  \bibinfo {journal} {Phys. Rev. Lett.}\ }%
  \textbf{\bibinfo {volume} {108}},\ \bibinfo {pages} {180601} (\bibinfo {year}
  {2012})\BibitemShut{NoStop}%
\bibitem{Mendl-Spohn-13}%
  \BibitemOpen
  \bibfield{author}{%
  \bibinfo {author} {\bibfnamefont{C.~B.}\ \bibnamefont{Mendl}}\ and\ \bibinfo
  {author} {\bibfnamefont{H.}~\bibnamefont{Spohn}},\ }%
  \bibfield{title}{%
  \enquote{\bibinfo {title} {Dynamic correlators of {F}ermi-{P}asta-{U}lam
  chains and nonlinear fluctuating hydrodynamics},}\ }%
  \bibfield{journal}{%
  \bibinfo {journal} {Phys. Rev. Lett.}\ }%
  \textbf{\bibinfo {volume} {111}},\ \bibinfo {pages} {230601} (\bibinfo {year}
  {2013})\BibitemShut{NoStop}%
\bibitem{Pikovsky-Shepelyansky-08}%
  \BibitemOpen
  \bibfield{author}{%
  \bibinfo {author} {\bibfnamefont{A.~S.}\ \bibnamefont{Pikovsky}}\ and\
  \bibinfo {author} {\bibfnamefont{D.~L.}\ \bibnamefont{Shepelyansky}},\ }%
  \bibfield{title}{%
  \enquote{\bibinfo {title} {Destruction of {A}nderson localization by a weak
  nonlinearity},}\ }%
  \bibfield{journal}{%
  \bibinfo {journal} {Phys. Rev. Lett.}\ }%
  \textbf{\bibinfo {volume} {100}},\ \bibinfo {pages} {094101} (\bibinfo {year}
  {2008})\BibitemShut{NoStop}%
\bibitem{Fishman-Krivolapov-Soffer-12}%
  \BibitemOpen
  \bibfield{author}{%
  \bibinfo {author} {\bibfnamefont{S.}~\bibnamefont{Fishman}}, \bibinfo
  {author} {\bibfnamefont{Y.}~\bibnamefont{Krivolapov}},\ and\ \bibinfo
  {author} {\bibfnamefont{A.}~\bibnamefont{Soffer}},\ }%
  \bibfield{title}{%
  \enquote{\bibinfo {title} {{The nonlinear {S}chr{\"o}dinger equation with a
  random potential: results and puzzles}},}\ }%
  \bibfield{journal}{%
  \bibinfo {journal} {Nonlinearity}\ }%
  \textbf{\bibinfo {volume} {25}},\ \bibinfo {pages} {R53--R72} (\bibinfo
  {year} {2012})\BibitemShut{NoStop}%
\bibitem{Lapteva_etal-14}%
  \BibitemOpen
  \bibfield{author}{%
  \bibinfo {author} {\bibfnamefont{T.~V.}\ \bibnamefont{Laptyeva}}, \bibinfo
  {author} {\bibfnamefont{M.~V.}\ \bibnamefont{Ivanchenko}},\ and\ \bibinfo
  {author} {\bibfnamefont{S.}~\bibnamefont{Flach}},\ }%
  \bibfield{title}{%
  \enquote{\bibinfo {title} {Nonlinear lattice waves in heterogeneous media},}\
  }%
  \bibfield{journal}{%
  \bibinfo {journal} {J. Phys. A}\ }%
  \textbf{\bibinfo {volume} {47}},\ \bibinfo {pages} {493001} (\bibinfo {year}
  {2014})\BibitemShut{NoStop}%
\bibitem{Skokos_etal-09}%
  \BibitemOpen
  \bibfield{author}{%
  \bibinfo {author} {\bibfnamefont{Ch.}\ \bibnamefont{Skokos}}, \bibinfo
  {author} {\bibfnamefont{D.~O.}\ \bibnamefont{Krimer}}, \bibinfo {author}
  {\bibfnamefont{S.}~\bibnamefont{Komineas}},\ and\ \bibinfo {author}
  {\bibfnamefont{S.}~\bibnamefont{Flach}},\ }%
  \bibfield{title}{%
  \enquote{\bibinfo {title} {{Delocalization of wave packets in disordered
  nonlinear chains}},}\ }%
  \bibfield{journal}{%
  \bibinfo {journal} {{Phys. Rev. E}}\ }%
  \textbf{\bibinfo {volume} {{79}}},\ \bibinfo {pages} {{056211}} (\bibinfo
  {year} {{2009}})\BibitemShut{NoStop}%
\bibitem{Mulansky-Ahnert-Pikovsky-11}%
  \BibitemOpen
  \bibfield{author}{%
  \bibinfo {author} {\bibfnamefont{M.}~\bibnamefont{Mulansky}}, \bibinfo
  {author} {\bibfnamefont{K.}~\bibnamefont{Ahnert}},\ and\ \bibinfo {author}
  {\bibfnamefont{A.}~\bibnamefont{Pikovsky}},\ }%
  \bibfield{title}{%
  \enquote{\bibinfo {title} {Scaling of energy spreading in strongly nonlinear
  disordered lattices},}\ }%
  \bibfield{journal}{%
  \bibinfo {journal} {Phys. Rev. E}\ }%
  \textbf{\bibinfo {volume} {83}},\ \bibinfo {pages} {026205} (\bibinfo {year}
  {2011})\BibitemShut{NoStop}%
\bibitem{Mulansky-Pikovsky-12a}%
  \BibitemOpen
  \bibfield{author}{%
  \bibinfo {author} {\bibfnamefont{M.}~\bibnamefont{Mulansky}}\ and\ \bibinfo
  {author} {\bibfnamefont{A.}~\bibnamefont{Pikovsky}},\ }%
  \bibfield{title}{%
  \enquote{\bibinfo {title} {Scaling properties of energy spreading in
  nonlinear {H}amiltonian two-dimensional lattices},}\ }%
  \bibfield{journal}{%
  \bibinfo {journal} {Phys. Rev. E}\ }%
  \textbf{\bibinfo {volume} {86}},\ \bibinfo {pages} {056214} (\bibinfo {year}
  {2012})\BibitemShut{NoStop}%
\bibitem{Achilleos-18}%
  \BibitemOpen
  \bibfield{author}{%
  \bibinfo {author} {\bibfnamefont{V.}~\bibnamefont{Achilleos}}, \bibinfo
  {author} {\bibfnamefont{G.}~\bibnamefont{Theocharis}},\ and\ \bibinfo
  {author} {\bibfnamefont{Ch.}\ \bibnamefont{Skokos}},\ }%
  \bibfield{title}{%
  \enquote{\bibinfo {title} {Chaos and {A}nderson-like localization in
  polydisperse granular chains},}\ }%
  \bibfield{journal}{%
  \bibinfo {journal} {Phys. Rev. E}\ }%
  \textbf{\bibinfo {volume} {97}},\ \bibinfo {pages} {042220} (\bibinfo {year}
  {2018})\BibitemShut{NoStop}%
\bibitem{Mulansky-Ahnert-Pikovsky-Shepelyansky-11}%
  \BibitemOpen
  \bibfield{author}{%
  \bibinfo {author} {\bibfnamefont{M.}~\bibnamefont{Mulansky}}, \bibinfo
  {author} {\bibfnamefont{K.}~\bibnamefont{Ahnert}}, \bibinfo {author}
  {\bibfnamefont{A.}~\bibnamefont{Pikovsky}},\ and\ \bibinfo {author}
  {\bibfnamefont{D.}~\bibnamefont{Shepelyansky}},\ }%
  \bibfield{title}{%
  \enquote{\bibinfo {title} {Strong and weak chaos in weakly nonintegrable
  many-body {H}amiltonian systems},}\ }%
  \bibfield{journal}{%
  \bibinfo {journal} {J. Stat. Phys.}\ }%
  \textbf{\bibinfo {volume} {145}},\ \bibinfo {pages} {1256--1274} (\bibinfo
  {year} {2011})\BibitemShut{NoStop}%
\bibitem{Pikovsky-Fishman-11}%
  \BibitemOpen
  \bibfield{author}{%
  \bibinfo {author} {\bibfnamefont{A.}~\bibnamefont{Pikovsky}}\ and\ \bibinfo
  {author} {\bibfnamefont{S.}~\bibnamefont{Fishman}},\ }%
  \bibfield{title}{%
  \enquote{\bibinfo {title} {Scaling properties of weak chaos in nonlinear
  disordered lattices},}\ }%
  \bibfield{journal}{%
  \bibinfo {journal} {Phys. Rev. E}\ }%
  \textbf{\bibinfo {volume} {83}},\ \bibinfo {pages} {025201(R)} (\bibinfo
  {year} {2011})\BibitemShut{NoStop}%
\bibitem{Senyange_etal-18}%
  \BibitemOpen
  \bibfield{author}{%
  \bibinfo {author} {\bibfnamefont{B.}~\bibnamefont{Senyange}}, \bibinfo
  {author} {\bibfnamefont{B.~Many}\ \bibnamefont{Manda}},\ and\ \bibinfo
  {author} {\bibfnamefont{Ch.}\ \bibnamefont{Skokos}},\ }%
  \bibfield{title}{%
  \enquote{\bibinfo {title} {Characteristics of chaos evolution in
  one-dimensional disordered nonlinear lattices},}\ }%
  \bibfield{journal}{%
  \bibinfo {journal} {Phys. Rev. E}\ }%
  \textbf{\bibinfo {volume} {98}},\ \bibinfo {pages} {052229} (\bibinfo {year}
  {2018})\BibitemShut{NoStop}%
\bibitem{Ngapasare_etal-19}%
  \BibitemOpen
  \bibfield{author}{%
  \bibinfo {author} {\bibfnamefont{A.}~\bibnamefont{Ngapasare}}, \bibinfo
  {author} {\bibfnamefont{G.}~\bibnamefont{Theocharis}}, \bibinfo {author}
  {\bibfnamefont{O.}~\bibnamefont{Richoux}}, \bibinfo {author}
  {\bibfnamefont{Ch.}\ \bibnamefont{Skokos}},\ and\ \bibinfo {author}
  {\bibfnamefont{V.}~\bibnamefont{Achilleos}},\ }%
  \bibfield{title}{%
  \enquote{\bibinfo {title} {Chaos and {A}nderson localization in disordered
  classical chains: {H}ertzian versus {F}ermi-{P}asta-{U}lam-{T}singou
  models},}\ }%
  \bibfield{journal}{%
  \bibinfo {journal} {Phys. Rev. E}\ }%
  \textbf{\bibinfo {volume} {99}},\ \bibinfo {pages} {032211} (\bibinfo {year}
  {2019})\BibitemShut{NoStop}%
\bibitem{Johansson-Kopidakis-Aubry-10}%
  \BibitemOpen
  \bibfield{author}{%
  \bibinfo {author} {\bibfnamefont{M.}~\bibnamefont{Johansson}}, \bibinfo
  {author} {\bibfnamefont{G.}~\bibnamefont{Kopidakis}},\ and\ \bibinfo {author}
  {\bibfnamefont{S.}~\bibnamefont{Aubry}},\ }%
  \bibfield{title}{%
  \enquote{\bibinfo {title} {{KAM} tori in {1D} random discrete nonlinear
  {S}chr{\"o}dinger model?}.}\ }%
  \bibfield{journal}{%
  \bibinfo {journal} {Europhys. Lett.}\ }%
  \textbf{\bibinfo {volume} {91}},\ \bibinfo {pages} {50001} (\bibinfo {year}
  {2010})\BibitemShut{NoStop}%
\bibitem{Prosen-Robnik-92}%
  \BibitemOpen
  \bibfield{author}{%
  \bibinfo {author} {\bibfnamefont{T.}~\bibnamefont{Prosen}}\ and\ \bibinfo
  {author} {\bibfnamefont{M.}~\bibnamefont{Robnik}},\ }%
  \bibfield{title}{%
  \enquote{\bibinfo {title} {Energy transport and detailed verification of
  fourier law in a chain of colliding harmonic oscillators},}\ }%
  \bibfield{journal}{%
  \bibinfo {journal} {J. Phys. A}\ }%
  \textbf{\bibinfo {volume} {25}},\ \bibinfo {pages} {3449--3472} (\bibinfo
  {year} {1992})\BibitemShut{NoStop}%
\bibitem{Roy-Pikovsky-12}%
  \BibitemOpen
  \bibfield{author}{%
  \bibinfo {author} {\bibfnamefont{S.}~\bibnamefont{Roy}}\ and\ \bibinfo
  {author} {\bibfnamefont{A.}~\bibnamefont{Pikovsky}},\ }%
  \bibfield{title}{%
  \enquote{\bibinfo {title} {Spreading of energy in the {D}ing-{D}ong model},}\
  }%
  \bibfield{journal}{%
  \bibinfo {journal} {CHAOS}\ }%
  \textbf{\bibinfo {volume} {22}},\ \bibinfo {eid} {026118} (\bibinfo {year}
  {2012})\BibitemShut{NoStop}%
\bibitem{Nesterenko-83}%
  \BibitemOpen
  \bibfield{author}{%
  \bibinfo {author} {\bibfnamefont{V.~F.}\ \bibnamefont{Nesterenko}},\ }%
  \bibfield{title}{%
  \enquote{\bibinfo {title} {Propagation of nonlinear compression pulses in
  granular media},}\ }%
  \bibfield{journal}{%
  \bibinfo {journal} {J. Appl. Mech. Tech. Phys.}\ }%
  \textbf{\bibinfo {volume} {24}},\ \bibinfo {pages} {733--743} (\bibinfo
  {year} {1983})\BibitemShut{NoStop}%
\bibitem{Rosenau-Hyman-93}%
  \BibitemOpen
  \bibfield{author}{%
  \bibinfo {author} {\bibfnamefont{P.}~\bibnamefont{Rosenau}}\ and\ \bibinfo
  {author} {\bibfnamefont{J.~M.}\ \bibnamefont{Hyman}},\ }%
  \bibfield{title}{%
  \enquote{\bibinfo {title} {Compactons: {S}olitons with finite wavelength},}\
  }%
  \bibfield{journal}{%
  \bibinfo {journal} {Phys. Rev. Lett.}\ }%
  \textbf{\bibinfo {volume} {70}},\ \bibinfo {pages} {564--567} (\bibinfo
  {year} {1993})\BibitemShut{NoStop}%
\bibitem{Rosenau-Pikovsky-05}%
  \BibitemOpen
  \bibfield{author}{%
  \bibinfo {author} {\bibfnamefont{P.}~\bibnamefont{Rosenau}}\ and\ \bibinfo
  {author} {\bibfnamefont{A.}~\bibnamefont{Pikovsky}},\ }%
  \bibfield{title}{%
  \enquote{\bibinfo {title} {Phase compactons in chains of dispersively coupled
  oscillators},}\ }%
  \bibfield{journal}{%
  \bibinfo {journal} {Phys. Rev. Lett.}\ }%
  \textbf{\bibinfo {volume} {94}},\ \bibinfo {pages} {174102} (\bibinfo {year}
  {2005})\BibitemShut{NoStop}%
\bibitem{Pikovsky-Rosenau-06}%
  \BibitemOpen
  \bibfield{author}{%
  \bibinfo {author} {\bibfnamefont{A.}~\bibnamefont{Pikovsky}}\ and\ \bibinfo
  {author} {\bibfnamefont{P.}~\bibnamefont{Rosenau}},\ }%
  \bibfield{title}{%
  \enquote{\bibinfo {title} {Phase compactons},}\ }%
  \bibfield{journal}{%
  \bibinfo {journal} {Physica D}\ }%
  \textbf{\bibinfo {volume} {218}},\ \bibinfo {pages} {56--69} (\bibinfo {year}
  {2006})\BibitemShut{NoStop}%
\bibitem{Ahnert-Pikovsky-09}%
  \BibitemOpen
  \bibfield{author}{%
  \bibinfo {author} {\bibfnamefont{K.}~\bibnamefont{Ahnert}}\ and\ \bibinfo
  {author} {\bibfnamefont{A.}~\bibnamefont{Pikovsky}},\ }%
  \bibfield{title}{%
  \enquote{\bibinfo {title} {Compactons and chaos in strongly nonlinear
  lattices},}\ }%
  \bibfield{journal}{%
  \bibinfo {journal} {Phys. Rev. E}\ }%
  \textbf{\bibinfo {volume} {79}},\ \bibinfo {pages} {026209} (\bibinfo {year}
  {2009})\BibitemShut{NoStop}%
\bibitem{Mulansky-Pikovsky-13}%
  \BibitemOpen
  \bibfield{author}{%
  \bibinfo {author} {\bibfnamefont{M.}~\bibnamefont{Mulansky}}\ and\ \bibinfo
  {author} {\bibfnamefont{A.}~\bibnamefont{Pikovsky}},\ }%
  \bibfield{title}{%
  \enquote{\bibinfo {title} {Energy spreading in strongly nonlinear disordered
  lattices},}\ }%
  \bibfield{journal}{%
  \bibinfo {journal} {New Journal of Physics}\ }%
  \textbf{\bibinfo {volume} {15}},\ \bibinfo {pages} {053015} (\bibinfo {year}
  {2013})\BibitemShut{NoStop}%
\end{thebibliography}
\end{document}